\documentclass[12pt, centerh1]{article}
\usepackage{natbib}
\usepackage[margin=1in]{geometry}
\usepackage{amsmath,natbib,multirow}
\usepackage{amsfonts,mathrsfs,moreverb,lipsum}
\usepackage{graphicx,colonequals}
\usepackage{fixltx2e}
\usepackage{color}
\usepackage{caption}
\usepackage{subcaption}
\usepackage{mathtools}
\usepackage{pbox}
\usepackage{booktabs}
\usepackage{dsfont}
\usepackage{authblk}

\newcommand{\btheta}{{\boldsymbol{\theta}}}
\newcommand{\bvtheta}{{\boldsymbol{\vartheta}}}
\newcommand{\bvphi}{{\boldsymbol{\varphi}}}

\newcommand{\vecx}{\mathbf{x}}

\newcommand{\vecX}{\mathbf{X}}

\newcommand{\vecv}{\mathbf{v}}
\newcommand{\vecz}{\mathbf{z}}

\newcommand{\matsig}{\mathbf\Sigma}
\newcommand{\matPsi}{\mathbf\Psi}

\newcommand{\tr}{\,\mbox{tr}}

\newcommand{\matm}{\mathbf{M}}

\newcommand{\vecc}{\text{vec}}
\newcommand{\fX}{\mathscr{X}}

\allowdisplaybreaks
\title{Mixtures of Contaminated Matrix Variate Normal Distributions}
\author[1]{Salvatore D. Tomarchio}
\author[2]{Michael P.B. Gallaugher}
\author[1]{Antonio Punzo}
\author[2]{Paul D. McNicholas}

\affil[1]{Department of Economics and Business, University of Catania, Catania, Italy}
\affil[2]{Department of Mathematics and Statistics, McMaster University, Ontario, Canada}

\begin{document}

\maketitle{}

\begin{abstract}

Analysis of three-way data is becoming ever more prevalent in the literature, especially in the area of clustering and classification.
Real data, including real three-way data, are often contaminated by potential outlying observations. 
Their detection, as well as the development of robust models insensitive to their presence, is particularly important for this type of data because of the practical issues concerning their effective visualization.
Herein, the contaminated matrix variate normal distribution is discussed and then utilized in the mixture model paradigm for clustering.
One key advantage of the proposed model is the ability to automatically detect potential outlying matrices by computing their \textit{a posteriori} probability to be a ``good'' or ``bad'' point. Such detection is currently unavailable using existing matrix variate methods.
An expectation conditional maximization algorithm is used for parameter estimation, and both simulated and real data are used for illustration.

\end{abstract}

\section{Introduction}
\label{sec:Int}
Nowadays there is an increased interest in the analysis of three-way (matrix variate) data, specifically in the area of clustering and classification via mixture models (see, e.g., recent contributions by \citealp{gallaugher18a,gallaugher20,melnykov2019studying, silva19,sarkar20}).
This data structure occurs from the observation of various attributes, measured on a set of units, in different situations or on different occasions.
Some typical examples include spatial multivariate data, multivariate longitudinal data and spatio-temporal data.
In all these cases we observe a matrix for each statistical unit, implying that a sample of $N$ random matrices can be arranged in a three-way array characterized by the following three dimensions: variables (rows), occasions (columns) and units (layers).

Real data, including three-way data, are quite often contaminated by outliers.
Outlier detection, as well as the development of robust models insensitive to their presence, constitute crucial problems. This is especially true for model-based clustering methods where just a few outliers can lead to severely biased estimates, incorrect classifications, and over fitting the number of groups.

Outliers, similar to atypical observations in general, may be roughly divided into two categories, mild and gross (see, \citealp{ritter15}, pp.\ 79--80 for details).
Herein we focus on mild outliers, which we also refer to as ``bad'' points following the nomenclature of \citet{aitkin80}. In the clustering context, mild outliers are points that deviate from the distribution within a cluster, but would fit well if the overall within-cluster distribution had heavier tails \citep{farcomeni2019}. The Gaussian distribution, although widely used in the literature, is often incapable of modelling data with mild outliers. The most common solution for managing this type of outlier is to fit a mixture of heavy-tailed elliptical distributions to the data.
In the multivariate literature, several models have been proposed to cope with this issue, such as mixtures of $t$ distributions \citep{peel00}, mixtures of power exponential distributions \citep{dang15} and mixtures of contaminated normal distributions \citep{punzo2016parsimonious}. 
Unfortunately, the corresponding three-way literature is far more limited. To our knowledge, the matrix variate $t$ distribution is the only symmetric matrix distribution with heavy tails used in the mixture model paradigm \citep{dougru16}.

In this paper we first discuss the contaminated matrix variate normal distribution including its useful properties and the interpretation of the parameters. This distribution is then used within the mixture model paradigm, generalizing the work of \citet{punzo2016parsimonious} to three-way data. One of the key advantages of this model when compared to the matrix $t$ distribution is the capability to automatically detect potential outliers by computing their \textit{a posteriori} probability to be ``good'' or ``bad'' points.

The remainder of this paper is laid out as follows.
Section~\ref{sec:Back} presents a detailed background and Section~\ref{sec:method} discusses the contaminated matrix variate normal distribution, its use in the mixture model setting and a variant of the classical EM algorithm for parameter estimation.
Furthermore, some notes on robustness are also provided, along with details on outlier detection and model performance evaluation.
A sensitivity analysis based on simulated data is presented in Section~\ref{sec:sens} and a real data application displaying the utility of the proposed method is discussed in Section~\ref{sec:real}.
This paper concludes with some conclusions and possible avenues for future work (Section~\ref{sec:conc}).

\section{Background}
\label{sec:Back}

\subsection{Finite Mixture Modelling and Clustering}
\label{sec:finite}

Clustering can be thought of as the process of finding and analyzing underlying group structure in heterogeneous data.
One common method for clustering is model-based clustering and makes use of a finite mixture model \citep[see][for extensive details]{mcnicholas16a}.
Let $\vecx_1,\ldots,\vecx_N$ be an observed sample consisting of $N$ $p$-variate realizations from a finite mixture model with probability density function (pdf)
\begin{equation}
f(\vecx~|~\bvtheta)=\sum_{g=1}^G\pi_gf_g(\vecx~|~\btheta_g).
\label{eq:mixtmodel}
\end{equation}
In~\eqref{eq:mixtmodel}, $\bvtheta$ denotes the overall parameter vector, $G$ represents the number of components, $f_g(\vecx~|~\btheta_g)$ is the $g$th mixture component with corresponding parameter $\btheta_g$ and weight (also known as mixing proportion) $\pi_g$, subject to conditions $\pi_g>0$ and $\sum_{g=1}^G\pi_g=1$. 

Due to its mathematical tractability, the Gaussian mixture model is frequently studied in the literature, with a history dating back to \cite{wolfe65}; however, as mentioned in Section~\ref{sec:Int}, issues arise when the data contain outlying observations. Specifically, using a mixture of Gaussians in the presence of mild outliers might result in over fitting the number of groups, and lead to severely biased parameter estimates.
For this reason, \cite{punzo2016parsimonious} present a mixture of contaminated normal distributions, which is able to model data with potential outliers and determine whether a specific point is a potential outlier with respect to a particular cluster. Other approaches provide heavy-tailed alternatives to the Gaussian mixture model, but do have the benefit of detecting outlying observations \citep{peel00, andrews11a,andrews11b,andrews12, lin14,dang15}.
Recent reviews of mixture model-based clustering are provided by \cite{bouveyron14} and \cite{mcnicholas16b}.

\subsection{Matrix Normal Distribution}
\label{sec:clust}

Similar to the multivariate case, the matrix normal distribution is most commonly used for clustering three-way data.
A random $r\times p$ matrix $\fX$ is said to follow a matrix variate normal distribution if its pdf can be written 
$$
f_{\text{MVN}}(\vecX~|~\matm,\matsig,\matPsi)=\frac{1}{(2\pi)^{\frac{rp}{2}}|\matsig|^{\frac{p}{2}}|\matPsi|^{\frac{r}{2}}}\exp\left\{-\frac{1}{2}\tr\left[\matsig^{-1}(\vecX-\matm)\matPsi^{-1}(\vecX-\matm)'\right]\right\},
$$
where $\matm$ is a mean matrix, $\matsig$ is an $r\times r$ row covariance matrix, and $\matPsi$ is a $p\times p$ column covariance matrix. One interesting property of the matrix normal distribution is its close relationship to the multivariate normal distribution via
\begin{equation}
\fX\sim \mathcal{N}_{r\times p}(\matm,\matsig,\matPsi) \iff \vecc(\fX)\sim \mathcal{N}_{rp}(\vecc(\matm),\matPsi\otimes \matsig),
\label{eq:normprop}
\end{equation}
where $\vecc(\cdot)$ is the vectorization operator and $\otimes$ is the Kronecker product.

The matrix variate normal distribution has the desirable feature of simultaneously modelling and identifying the between and within-variable variabilities as well as reducing the number of free parameters from $rp(rp + 1)/2$ to $r(r + 1)/2 + p(p + 1)/2$. 
It is important to note that the matrix variate normal distribution is not identifiable because the matrices $\matsig$ and $\matPsi$ are only unique up to a positive constant; however, their Kronecker product $\matPsi\otimes \matsig$ is uniquely defined. Herein, to resolve the identifiability problem, we set the first diagonal element of $\matsig$ equal to one.

\section{Methodology}
\label{sec:method}

\subsection{Contaminated Matrix Normal Distribution}
\label{sec:cn}

An $r\times p$ random matrix $\fX$ is said to follow a contaminated matrix variate normal (CMVN) distribution, if its density can be written
\begin{equation}
f_{\text{CMVN}}(\vecX~|~\matm,\matsig,\matPsi,\eta,\alpha)=\alpha f_{\text{MVN}} (\vecX~|~\matm,\matsig,\matPsi)+(1-\alpha) f_{\text{MVN}} (\vecX~|~\matm,\eta\matsig,\matPsi),
\label{eq:mixtcontmvx}
\end{equation}
with $0<\alpha<0.5$ and $\eta>1$.
This distribution belongs to the absolutely continuous elliptically contoured distributions family \citep{gupta1994} and, similar to the multivariate case, it is represented in the form of a two-component mixture model. 
The first component, with mixing proportion $\alpha$, models the points that are considered ``good".
The second component, which inflates the variance with the weight $\eta$, models the outlying observations that are considered ``bad".
Therefore, a useful characteristic of this distribution is the practical interpretation of its parameters, with $\alpha$ being the proportion of ``good'' matrices and $\eta$ denoting the degree of contamination. The degree of contamination can be interpreted as an inflation parameter and is a measure of how different the outlying matrices are from the bulk of the good data.

Another important characteristic of this distribution is that once the parameters are estimated, it is possible to determine whether a generic matrix, say $\vecX_i$, is good via the maximum {\it a posteriori} probability
\begin{equation}
\hat{v}_{i}\colonequals\frac{\hat{\alpha }f_{\text{MVN}} (\vecX_i~|~\hat{\matm},\hat{\matsig},\hat{\matPsi})}{f_{\text{CMVN}}(\vecX_i~|~\hat{\matm},\hat{\matsig},\hat{\matPsi},\hat{\eta},\hat{\alpha})}.
\label{eq:probability good}
\end{equation}
Specifically, $\vecX_i$ will be considered good if $\hat{v}_{i}>0.5$, while it will be considered bad otherwise.
This aspect is of particular importance for three-way data given that visualization techniques --- and, therefore, the visual detection of outlying matrices --- is a challenging task.

\subsection{Remark on the inflation parameter}
\label{sec:note}
As mentioned in Section~\ref{sec:clust}, the scale matrices $\matsig$ and $\matPsi$ are unique only up to a positive multiplicative constant and the Kronecker product $\matPsi\otimes \matsig$ is uniquely defined.
Moreover, if $\sigma_{jj}$ and $\psi_{ll}$ are the $j$th and $l$th diagonal elements of $\matsig$ and $\matPsi$, respectively, then the variance of the element $jl$ in the matrix $\fX$ is $\sigma_{jj}\psi_{ll}$, a result easily obtained from \eqref{eq:normprop}.
Therefore, to increase the overall variance to model the ``bad matrices", a weight $\eta$ need only be applied to one of the scale matrices.
In fact, if weights  $\eta_{\Sigma}$ and $\eta_{\Psi}$ are applied to $\matsig$ and $\matPsi$ respectively, then
$$
(\eta_{\Psi}\matPsi)\otimes(\eta_{\Sigma}\matsig)=\eta_{\Psi}\eta_{\Sigma}(\matPsi\otimes\matsig)=\eta(\matPsi\otimes\matsig)=\matPsi\otimes(\eta\matsig),
$$
where $\eta=\eta_{\Psi}\eta_{\Sigma}$.

\subsection{Mixtures of CMVN Distributions}
\label{sec:mixtCN}

The CMVN distribution is now considered in the mixture model context for its use in clustering and classification.
Specifically, an $r\times p$ random matrix $\fX$ with realization $\vecX$ has density 
$$
f(\vecX~|~\bvtheta)=\sum_{g=1}^G \pi_g f_{\text{CMVN}}(\vecX~|~\matm_g,\matsig_g,\matPsi_g,\eta_g,\alpha_g).
$$
To find maximum likelihood (ML) estimates for the parameters of our model, we adopt the expectation conditional maximization (ECM) algorithm \citep{meng93}.
The ECM algorithm is a variant of the classical expectation-maximization (EM) algorithm \citep{dempster77}, which is a natural approach for ML estimation when data are incomplete.
The two sources of missingness in this case are:
\begin{itemize} \item the unknown group memberships $\vecz_1,\ldots,\vecz_N$, where $\vecz_i=(z_{i1},\dots,z_{iG})'$ so that $z_{ig}=1$ if observation $i$ is in group $g$ and $z_{ig}=0$ otherwise; and \item the classification of observation $i$ in group $g$ as either good or bad, i.e.~$\vecv_1,\ldots,\vecv_N$, where $\vecv_i=(v_{i1},\dots,v_{iG})'$ so that $v_{ig}=1$ if observation $i$ in group $g$ is good and $v_{ig}=0$, otherwise.
\end{itemize}

The complete-data is therefore given by $\left\{\vecX_1,\ldots,\vecX_N,\vecz_1,\ldots,\vecz_N,\vecv_1,\ldots,\vecv_N\right\}$ and the complete-data log-likelihood can be written
$$
\ell_C(\boldsymbol \theta)=\ell_{C1}({\boldsymbol \pi})+\ell_{C2}({\boldsymbol \alpha})+\ell_{C3}({\bvphi}),
$$
where 
$$
\ell_{C1}({\boldsymbol \pi})=\sum_{i=1}^N\sum_{g=1}^Gz_{ig}\log\pi_g,
$$
$$
\ell_{C2}({\boldsymbol \alpha})=\sum_{i=1}^N\sum_{g=1}^Gz_{ig}[v_{ig}\log\alpha_g+(1-v_{ig})\log(1-\alpha_g)],
$$
and 
\begin{equation}
\begin{aligned}
\ell_{C3}(\bvphi)&=\sum_{i=1}^N\sum_{g=1}^Gz_{ig} \Big[ -\frac{p}{2}\log|\matsig_g|-\frac{r}{2}\log|\matPsi_g|-\frac{rp}{2}(1-v_{ig})\log(\eta_g)\\
&\qquad\qquad\qquad -\frac{1}{2}\left(v_{ig}+\frac{1-v_{ig}}{\eta_g}\right)\tr\left\{\matsig_g^{-1}(\vecX_i-\matm_g)\matPsi_g^{-1}(\vecX_i-\matm_g)'\right\}\Big],
\end{aligned}
\end{equation}
with $\boldsymbol \pi =(\pi_1,\ldots,\pi_G)'$, $\boldsymbol \alpha =(\alpha_1,\ldots,\alpha_G)'$ and $\bvphi = \left\{\matm_g,\matsig_g,\matPsi_g,\eta_g\right\}$ for $g\in\{1,\ldots,G\}$.
After initialization the ECM algorithm proceeds as follows where, following the notation of \citet{melnykov2019studying}, the parameters marked with one dot correspond to the previous iteration and those marked with two dots represent the updates at the current iteration.

\noindent{\bf E-Step}: Update $z_{ig}$ and $v_{ig}$ via
\begin{equation}
\begin{split}
\ddot{z}_{ig}&\colonequals\mathbb{E_\bvtheta}(Z_{ig}~|~\vecX_{i})=\frac{\dot{\pi}_g f_{\text{CMVN}}(\vecX_i~|~\dot{\matm}_g,\dot{\matsig}_g,\dot{\matPsi}_g,\dot{\eta}_g,\dot{\alpha}_g)}{\sum_{h=1}^G\dot{\pi}_hf_{\text{CMVN}}(\vecX_i~|~\dot{\matm}_h,\dot{\matsig}_h,\dot{\matPsi}_h,\dot{\eta}_h,\dot{\alpha}_h)},
\end{split}
\label{eq:E-step_Z}
\end{equation}
\begin{equation}
\begin{split}
\ddot{v}_{ig}&\colonequals\mathbb{E_\bvtheta}(V_{ig}~|~\vecX_{i})=\frac{\dot{\alpha}_g f_{\text{MVN}} (\vecX_i~|~\dot{\matm}_g,\dot{\matsig}_g,\dot{\matPsi}_g)}{f_{\text{CMVN}}(\vecX_i~|~\dot{\matm}_g,\dot{\matsig}_g,\dot{\matPsi}_g,\dot{\eta}_g,\dot{\alpha}_g)}.
\end{split}
\label{eq:E-step_V}
\end{equation}

\noindent{\bf CM-Step 1}: Update $\pi_g,\alpha_g,\matm_g$ according to
\begin{equation}
\ddot{\pi}_g=\frac{\ddot{N_g}}{N},\quad\ddot{\alpha}_g=\frac{\sum_{i=1}^N\ddot{z}_{ig}\ddot{v}_{ig}}{\ddot{N_g}},
\label{eq:M_step_pietc}
\end{equation}
\begin{equation}
\ddot{\matm}_g=\frac{1}{\ddot{s_g}}\sum_{i=1}^N\ddot{z}_{ig}\left(\ddot{v}_{ig}+\frac{1-\ddot{v}_{ig}}{\dot{\eta}_g}\right)\vecX_i, 
\label{eq:M_step_M}
\end{equation}
where $\ddot{N_g}=\sum_{i=1}^N\ddot{z}_{ig}$ and $\ddot{s_g}=\sum_{i=1}^N\ddot{z}_{ig}\left(\ddot{v}_{ig}+\frac{1-\ddot{v}_{ig}}{\dot{\eta}_g}\right)$. 

\noindent{\bf CM-Step 2}: Update $\matsig_g$ by
\begin{equation}
\ddot{\matsig}_g=\frac{1}{p\ddot{N_g}}\sum_{i=1}^N\ddot{z}_{ig}\left(\ddot{v}_{ig}+\frac{1-\ddot{v}_{ig}}{\dot{\eta}_g}\right)(\vecX_i-\ddot{\matm}_g)\dot{\matPsi}_g^{-1}(\vecX_i-\ddot{\matm}_g)'.
\label{eq:M_step_Sigma}
\end{equation}

\noindent{\bf CM-Step 3}: Update $\matPsi_g$ according to
\begin{equation}
\ddot{\matPsi}_g=\frac{1}{r\ddot{N_g}}\sum_{i=1}^N\ddot{z}_{ig}\left(\ddot{v}_{ig}+\frac{1-\ddot{v}_{ig}}{\dot{\eta}_g}\right)(\vecX_i-\ddot{\matm}_g)'\ddot{\matsig}_g^{-1}(\vecX_i-\ddot{\matm}_g).
\label{eq:M_step_Psi}
\end{equation}

\noindent{\bf CM-Step 4}: Update $\eta_g$ via
$$
\ddot{\eta}_g=\max\left\{\eta_{\text{min}},\frac{\sum_{i=1}^N\ddot{z}_{ig}(1-\ddot{v}_{ig})\tr[\ddot{\matsig}_g^{-1}(\vecX_i-\ddot{\matm}_g)\ddot{\matPsi}_{g}^{-1}(\vecX_i-\ddot{\matm}_g)']}{\sum_{i=1}^N\ddot{z}_{ig}(1-\ddot{v}_{ig})}\right\},
$$
where $\eta_{\text{min}}>1$.
In our analyses, we set $\eta_{\text{min}}=1.0001$.
Similarly to \citet{punzo2016parsimonious}, we start our ECM algorithm by randomly initializing the quantities involved in the E-step.

\subsection{Some notes on robustness}
\label{sec:robust}

Based on~\eqref{eq:M_step_M}, the update for $\matm_g$ is a weighted mean of the $\vecX_i$ values, with weights
\begin{equation}
v_{ig}+\frac{1-v_{ig}}{\eta_g}.
\label{eq:weights}
\end{equation}
Consider now the update for $v_{ig}$ in~\eqref{eq:E-step_V} as a function of the squared Mahalanobis distance $\delta_{ig}=\tr[\matsig_g^{-1}(\vecX_i-\matm_g)\matPsi_{g}^{-1}(\vecX_i-\matm_g)']$, i.e.,
\begin{align}
h\left(\delta_{ig};\alpha_{g},\eta_{g}\right)&=\alpha_{g}\exp\left(-\frac{\delta_{ig}}{2}\right)\left[\alpha_{g}\exp\left(-\frac{\delta_{ig}}{2}\right)+\frac{\left(1-\alpha_{g}\right)}{\sqrt{\eta^{rp}_{g}}}\exp\left(-\frac{\delta_{ig}}{2\eta_{g}}\right)\right]^{-1}\nonumber \\&=\left\{1+\frac{\left(1-\alpha_{g}\right)}{\alpha_{g}}\frac{1}{\sqrt{\eta^{rp}_{g}}}\exp\left[\frac{\delta_{ig}}{2}\left(1-\frac{1}{\eta_{g}}\right)\right]\right\}^{-1},
\label{eq:rob1}
\end{align}

Due to the constraint $\eta_{g}>1$, from~\eqref{eq:rob1} it is straightforward to realize that $h\left(\delta_{ig};\alpha_{g},\eta_{g}\right)$ is a decreasing function of $\delta_{ig}$.
Based on~\eqref{eq:rob1},~\eqref{eq:weights} can be written
\begin{equation}
w\left(\delta_{ig};\alpha_{g},\eta_{g}\right)=h\left(\delta_{ig};\alpha_{g},\eta_{g}\right)+\frac{1-h\left(\delta_{ig};\alpha_{g},\eta_{g}\right)}{\eta_{g}}=\frac{1}{\eta_{g}}\left[1+\left(\eta_{g}-1\right)h\left(\delta_{ig};\alpha_{g},\eta_{g}\right)\right].
\label{eq:w_wei}
\end{equation}
From \eqref{eq:w_wei}, it is easy to see that $w\left(\delta_{ig};\alpha_{g},\eta_{g}\right)$  is an increasing function of
$h\left(\delta_{ig};\alpha_{g},\eta_{g}\right)$; this also means that $w\left(\delta_{ig};\alpha_{g},\eta_{g}\right)$ is a decreasing function of $\delta_{ig}$.
Therefore, the weights in~\eqref{eq:weights} reduce the impact of bad points in the estimation of the means $\matm_g$, thereby providing robust estimates of these means.
Similarly, from~\eqref{eq:M_step_Sigma} and~\eqref{eq:M_step_Psi}, the larger $\delta_{ig}$ values also have smaller effect on $\matsig_g$ and $\matPsi_g$, $g=1,\ldots,G$, due to the weights in~\eqref{eq:weights}.

\subsection{Detection of bad matrices}
\label{sec:auto} 

For each matrix $\vecX_i$, once the ECM algorithm reaches convergence, we can determine both
its cluster of membership and whether it is a good or a bad matrix in that cluster.
Let $\hat{z}_{ig}$ and $\hat{v}_{ig}$ be the values at convergence of~\eqref{eq:E-step_Z} and~\eqref{eq:E-step_V}, respectively. 
The matrix $\vecX_i$ is assigned to the cluster maximizing the estimated {\it a posteriori} probabilities $\hat{z}_{ig}$.
We then consider $\vecX_i$ good in that cluster if $\hat{v}_{ih}> 0.5$, and bad otherwise.

\subsection{Model selection and performance assessment}
\label{sec:modelsec}

It is often the case that the number of groups $G$ is not known {\it a priori}, and model selection is commonly performed by computing a suitable (likelihood-based) model selection criterion.
The Bayesian information criterion \cite[BIC;][]{schwarz78} is one of the most popular, and will be used in the following analyses.
It is defined as:
$$\text{BIC}= 2\ell(\hat{\bvtheta})-m\log{N},$$
where $m$ is the overall number of free parameters in the model.
Note that, with this formulation, models with higher BIC values are preferred.

To assess classification performance, the adjusted rand index \cite[ARI;][]{hubert85} is used. The ARI evaluates the agreement between the true classification and the one predicted by the model.
An ARI of 1 indicates perfect agreement between the two partitions, while the expected value of the ARI under random classification is 0. Extensive details on the ARI are given by \cite{steinley04}.
In addition to the ARI, the misclassification rate (MCR), which is the proportion of units that are misclassified, will be used to assess classification performance.

\section{Sensitivity analysis}
\label{sec:sens}

\subsection{Overview}
A sensitivity study is here described to illustrate the behaviour of our model in the presence of bad points.
Specifically, in Section~\ref{sec:single} the impact of a single atypical observation on the fitting of CMVN mixtures is evaluated, while their performance in the presence of uniform noise is analyzed in Section~\ref{sec:noise}.
Both studies are based on an artificial dataset of size $N=150$, randomly generated from a mixture of two matrix variate normal distributions with $r=2$, $p=4$ and parameters displayed in Table~\ref{tab:param}. 
Therefore, each data point is a $2 \times 4$ matrix.
For comparison purposes, matrix variate $t$ (MVT) mixtures and matrix variate normal (MVN) mixtures are considered in the analyses herein.
\begin{table}[!ht]
\caption{Parameters used to generate the artificial dataset.} 
\centering
\begin{tabular*}{\textwidth}{@{\extracolsep{\fill}}lcc} 
\hline
Parameter & Group 1 & Group 2 \\
\hline 
	$\pi_{g}$ & 0.50 & 0.50  \\
	
  $\matm_{g}$   & $\setlength\arraycolsep{2.8pt}\begin{pmatrix} -2.60 & -1.10 & -0.50 & -0.20 \\
                                                                 1.30 &  0.60 &  0.30 &  0.10 \end{pmatrix}$   
						    & $\setlength\arraycolsep{2.8pt}\begin{pmatrix}  1.50 & 1.70 & 1.90 & 2.20 \\
			                                                          -3.70 &-2.70 &-2.00 &-1.50 \end{pmatrix}$\\
																																			
	$\matsig_{g}$ & $\setlength\arraycolsep{2.8pt}\begin{pmatrix}  2.00 & 0.00 \\
                                                                 0.00 & 1.00 \end{pmatrix}$ 
						    &	$\setlength\arraycolsep{2.8pt}\begin{pmatrix}  1.70 & 0.50 \\ 
										                                             0.50 & 1.30 \end{pmatrix}$\\
								      																					
	$\matPsi_{g}$ & $\setlength\arraycolsep{2.8pt}\begin{pmatrix}  1.00 & 0.50 & 0.25 & 0.13 \\
                                                                 0.50 & 1.00 & 0.50 & 0.25 \\
                                                                 0.25 & 0.50 & 1.00 & 0.50 \\  
																		                             0.13 & 0.25 & 0.50 & 1.00 \end{pmatrix}$ 
						    & $\setlength\arraycolsep{2.8pt}\begin{pmatrix}  1.00 & 0.50 & 0.25 & 0.13 \\
                                                                 0.50 & 1.00 & 0.50 & 0.25 \\
                                                                 0.25 & 0.50 & 1.00 & 0.50 \\  
																		                             0.13 & 0.25 & 0.50 & 1.00 \end{pmatrix}$ \\
\hline
\end{tabular*}
\label{tab:param} 
\end{table}

\subsection{Assessing the impact of a single atypical point}
\label{sec:single}

Ten ``perturbed'' versions of this dataset are created by adding to the sixth observation the matrix $c \mathds{1}$, where $\mathds{1}$ is a matrix of ones and  $c\in\left\{2,4,6,8,10,12,14,16,18,20\right\}$.
On every ``perturbed'' dataset, CMVN mixtures are fitted for $G\in\left\{1,2,3\right\}$.

In all of the considered cases, the BIC selects the true number of groups ($G=2$) when fitting the CMVN mixture model.
It is interesting to note that the fitted model detects the perturbed observation as a bad point for $c>2$, and it is the only bad point detected in each case.
The estimated {\it a posteriori} probabilities $\hat{v}_{6g}$ for the perturbed observation to be a good point and the estimated values of $\eta_g$ are shown in Table~\ref{tab:post}.
As we can see, the farther it is from the bulk of the data, the lower its probability of being a good point. In addition, this probability is practically null for $c>2$. Regarding the values of $\eta_g$, the more the perturbed observation departs from the bulk of the data, the higher the value of $\eta_g$, confirming its useful interpretation as an inflation parameter.
\begin{table}[!ht]
\centering
\caption{Estimated {\it a posteriori} probability $\hat{v}_{6g}$ of being a good point for each perturbed dataset.}%
\begin{tabular*}{0.8\textwidth}{@{\extracolsep{\fill}}llcc|lcc} 
\hline
$c$ & $\hat{v}_{6g}$&$\hat{\eta}_{g}$&&$c$ & $\hat{v}_{6g}$&$\hat{\eta}_g$ \\
\hline
2  & $9.9889 \times 10^{-01}$&1.01&&12 & $1.4419 \times 10^{-59}$&39.58 \\
4  & $7.7778 \times 10^{-05}$&3.44&&14 & $1.0673 \times 10^{-80}$&52.30 \\
6  & $4.0366 \times 10^{-13}$&6.38&&16 & $2.8339 \times 10^{-105}$ &66.43\\
8  & $4.1487 \times 10^{-27}$ &19.60&&18 & $2.3217 \times 10^{-132}$&82.91\\
10 & $1.1401 \times 10^{-41}$ &28.69&&20 & $2.7058 \times 10^{-163}$&100.24\\
\hline 
\end{tabular*}
 \label{tab:post}
 \end{table}

If an MVN mixture is fitted to the simulated perturbed data, the number of groups detected by the BIC becomes $G=3$ for $c>6$. On the other hand, although the MVT mixture performs similarly to the CMVN, the CMVN mixture both accounts for outliers and allows for their automatic identification, whereas the MVT mixture just accounts for the outliers.

From the analysis of the BIC values in Table~\ref{tab:BICsingle}, it is interesting to note that, for increased levels of contamination the best model chosen by the BIC gradually shifts from the MVN mixture to the CMVN mixture. Therefore, for those situations with a higher level of contamination, CMVN mixtures appear to provide a better fit of the data.
\begin{table}[!ht]
\centering
\caption{Groups chosen by the BIC, with corresponding values, for the competing mixture models on the perturbed datasets.} 
\begin{tabular*}{0.8\textwidth}{@{\extracolsep{\fill}}ccccccccccc} 
\hline
$c$ && \multicolumn{2}{c}{MVN}  && \multicolumn{2}{c}{MVT} && \multicolumn{2}{c}{CMVN} \\
\cline{3-4}\cline{6-7}\cline{9-10}
&&$G$&BIC&&$G$&BIC&&$G$&BIC \\
\hline
2 && 2 & \bfseries $-3955.38$ && 2 & $-3972.21$ && 2 & $-3985.44$ \\
4 && 2 & \bfseries $-3979.88$ && 2 & $-3988.14$ && 2 & $-3993.43$ \\
6 && 2 & $-4013.61$ && 2 & \bfseries $-4000.99$ && 2 & $-4004.76$ \\
8 && 3 & $-4036.85$ && 2 & \bfseries $-4010.32$ && 2 & $-4011.19$ \\
10 && 3 & $-4029.90$ && 2 & $-4017.32$ && 2 & \bfseries $-4014.10$ \\
12 && 3 & $-4034.00$ && 2 & $-4022.99$ && 2 & \bfseries $-4016.60 $\\
14 && 3 & $-4041.59$ && 2 & $-4027.48$ && 2 & \bfseries $-4018.78$ \\
16 && 3 & $-4054.99$ && 2 & $-4031.43$ && 2 & \bfseries $-4020.70 $\\
18 && 3 & $-4056.78$ && 2 & $-4036.66$ && 2 & \bfseries $-4022.43 $\\
20 && 3 & $-4043.12$ && 2 & $-4037.97$ && 2 & \bfseries $-4024.00$ \\ 
\hline
\end{tabular*}
\label{tab:BICsingle} 
\end{table}

\subsection{Assessing the impact of background noise}
\label{sec:noise}

In this application, $10\%$ of the points are randomly substituted by noisy matrices whose elements are generated from a uniform distribution over the interval $[-8, 8]$.
All the competing mixture models are fitted to the data with $G\in\left\{1,2,3\right\}$ and their results are shown in Table~\ref{tab:noise}.
Similar to \cite{punzo2016parsimonious}, the ARI and the misclassification rates are computed only with respect to the true good observations, i.e., by excluding the noisy points.
\begin{table}[!ht]
\caption{BIC values and classification performance for the MVN, MVT, and CMVN mixture models for the simulated data with uniform noise.} 
\centering
\begin{tabular*}{0.8\textwidth}{@{\extracolsep{\fill}}lcccc} 
\hline
 & $G$ & BIC & ARI & MCR \\
\hline
MVN   & 3 & $-4436.89$ & 0.98 & 0.75\% \\
MVT & 2 & $-4429.85$ & 0.94 & 1.48\% \\
CMVN  & 2 & \bfseries $-4396.69$ & \bfseries 1.00 & \bfseries 0.00\% \\
\hline
\end{tabular*}
\label{tab:noise} 
\end{table}

As in the previous simulation, the MVN mixture is affected by atypical observations.
The BIC selects an additional third component that is attempting to model part of the background noise. When fitting the MVT mixture, the correct number of groups is found by the BIC, but the resulting misclassification rate is worse than the MVN mixture. 

The use of the CMVN mixture results in the correct selection of the number of groups, and perfect classification of the good points.
Furthermore, the noisy observations are correctly classified as bad points, with estimated posterior probabilities to be good in the range $\left[3.0010 \times 10^{-57}, 5.3852 \times 10^{-11}\right]$.

\section{ANVUR data}
\label{sec:real}

The Italian National Agency for the Evaluation of Universities and Research Institutes (ANVUR) maintains data on Italian universities' quantitative indicators concerning the academic careers of the students as well as the results of their teaching activities. Such data are now considered.
In this application, the following three variables are measured over three years for $N=75$ study programs in the non-telematic Italian universities.
The three variables are the percentage of students that have earned at least 40 course credits during the calendar year, the percentage of students that continued in the second year of the same study program, and the percentage of students who would enrol again in the same course of study.
Each data point is then a $3 \times 3 $ matrix.
Every study program is measured at the national level, i.e., it is the average value of all the study programs of the same type across the country for the reference year.

There are $G=2$ groups in the data with $N_{1}=33$ bachelor's degrees and $N_{2}=42$ master's degrees.
The MVN, MVT, and CMVN mixtures are fitted to the data for $G\in\left\{1,2,3\right\}$ and their results are shown in Table~\ref{tab:anvur}. The BIC selects $G=3$ groups when fitting the MVN mixture, and the correct number of groups ($G=2$) when fitting both the MVT and CMVN mixtures.
In terms of model fit, fitting the CMVN mixture results in the highest BIC. Moreover, the best classification performance is achieved by fitting the CMVN mixture.
\begin{table}[!ht]
\caption{BIC values and classification performance of the MVN, MVT, and CMVN mixture models for the ANVUR data.} 
\centering
\begin{tabular*}{0.8\textwidth}{@{\extracolsep{\fill}}lcccc} 
\hline
 & $G$ & BIC & ARI & MCR \\
\hline
MVN   & 3 &   414.87 & 0.84 & 8.00\% \\
MVT & 2 &   465.01 & 0.84 & 4.00\% \\
CMVN  & 2 & \bfseries 469.11 & \bfseries 0.90 & \bfseries 2.66\% \\
\hline
\end{tabular*}
\label{tab:anvur} 
\end{table}

The estimated proportion of good points and the degree of contamination for the first group are $\hat{\alpha}_{1}=0.76$ and $\hat{\eta}_{1}=6.69$, respectively, whereas for the second group they are $\hat{\alpha}_{2}=0.98$ and $\hat{\eta}_{2}=63.74$.
Therefore, while there are more outliers found in group 1, the single outlier found in group 2 is more severe, as reflected by the much greater value of the inflation parameter.
These aspects can be better understood by looking at Table~\ref{tab:anvur_bad}, where the study programs marked as bad for the first group are reported, along with their estimated probabilities to be good points.
\begin{table}[!ht]
\caption{Study programs marked as bad in the first group using the CMVN mixture with corresponding $\hat{v}_{i1}$ values.} 
\centering
\begin{tabular*}{\textwidth}{@{\extracolsep{\fill}}lc} 
\hline
Study Program & $\hat{v}_{i1}$ \\
\hline
Territorial, Urban, Landscape and Environmental Planning Sciences  & $ 1.2231 \times 10^{-05}$ \\
Sciences and Technologies for the Environment and Nature  & $ 2.5089 \times 10^{-03}$ \\
Geological Sciences & $ 7.6853 \times 10^{-03}$\\
Social Service  & $ 5.1102 \times 10^{-16}$ \\
Sociology & $ 1.2060 \times 10^{-11}$ \\
Pharmaceutical Sciences and Technologies & $ 3.3421 \times 10^{-07}$ \\
Sciences and Techniques of Preventive and Adapted Physical Activities & $ 1.9402 \times 10^{-09}$ \\
Health Professions of Rehabilitation Sciences &  $ 2.1618 \times 10^{-08}$ \\
\hline
\end{tabular*}
\label{tab:anvur_bad} 
\end{table}

It is interesting to note that the first three study programs marked as bad in the first group are all related, in some way, to the natural and environmental sciences.
Similarly, the social services and sociology programs are closely related, and they are the only programs in this dataset that explicitly deal with this area.
The only study program flagged as being bad for the second group is the geophysical sciences program. For this observation, the probability of being a good point is very small ($4.2215 \times 10^{-114}$). 

\section{Conclusions}
\label{sec:conc}

A mixture of contaminated matrix variate normal distributions has been introduced and with its many useful properties have been discussed. The heavier tails of the CMVN distribution allow for the modelling of matrix variate data with outlying observations as well as reducing the impact of the outlying matrices on the parameter estimates. The most useful aspect of this model, however, is its ability to identify outlying matrices in a straightforward manner. This is very important in the analysis of three-way data because it is difficult to visualize the data.

These interesting aspects have been demonstrated both in a simulation study and in a real data application, where the contaminated matrix normal mixture model obtained better performance than the competing models in terms of model fit, classification, and the inherent ability to detect outlying observations. 

An interesting point for further research could be to accommodate asymmetric contamination by using skewed matrix variate distributions \citep{gallaugher17a, gallaugher19}. Another avenue, and one that is presently being considered, is the extension of the approach of \cite{clark19} to three-way data.


\end{document}